\documentclass[floatfix,aps,prl,twocolumn,showpacs,preprintnumbers,amsmath,amssymb]{revtex4}
\usepackage{epsfig}
\usepackage{graphicx}
\usepackage{dcolumn}
\usepackage{bm}
\usepackage{ulem}
\usepackage{color}

\begin{document}
\title{Superconductor--Insulator Transition in Long MoGe Nanowires}
\author{Hyunjeong Kim, Shirin Jamali, and A. Rogachev}
\affiliation{Department of Physics and Astronomy, University of
Utah, Salt Lake City, Utah 84112, USA}
\date{\today}

\begin {abstract}
Properties of one-dimensional superconducting wires depend on
physical processes with different characteristic lengths. To
identify the process dominant in the critical regime we have studied
transport properties of very narrow (9-20 nm) MoGe wires fabricated
by advanced electron-beam lithography in wide range of lengths, 1-25
$\mu$m. We observed that the wires undergo a
superconductor--insulator transition that is controlled by cross
sectional area of a wire and possibly also by  the
thickness-to-width ratio. Mean-field critical temperature decreases
exponentially with the inverse of the wire cross section. We
observed that qualitatively similar superconductor--insulator
transition can be induced by external magnetic field. Some of our
long superconducting MoGe nanowires can be identified as localized
superconductors, namely in these wires one-electron localization
length is much shorter than the length of a wire.
\end {abstract}


\maketitle

One-dimensional systems play a special role in physics since they
often allow a more simple theoretical description than their
counterparts in higher dimensions \cite{Giamarchi}. Moreover,
experimental testing of 1D systems with finite length can probe the
length scale of distinct physical processes. This possibility is
particularly important for systems that have long-range coherence in
3D, such as superconductors.
    In a one-dimensional limit superconductivity can be suppressed by
several processes. If a wire is microscopically disordered, enhanced
Coulomb repulsion competes with the Cooper pairing and suppresses an
amplitude of order parameter \cite{Oreg}. In a homogeneous wire this
process determines mean-field critical temperature. Below this
temperature, a superconducting wire can acquire resistance as a
result of phase slips (PS), topological fluctuations of the order
parameter. A phase slip can occur as a result of a
thermally-activated fluctuation (TAPS) \cite{Tinkham} or a quantum
fluctuation (QPS) \cite{AGZ}. The activation barrier for both
processes is the same and depends on the amplitude of the order
parameter and coherence length. Recent theories suggest that the QPS
rate can also depend on some long-scale (or external) parameters.
For example the QPS rate can be suppressed both in wires shorter
than the length of the phase propagation during a phase slip
\cite{Meidan} and in very long wires as a result of attractive
interaction between QPS with different signs \cite{GZ_PRL}. The
quantum state of a wire is also predicted to depend on the state
\cite{Sachdev} and impedance \cite{Khlebnikov} of macroscopic
electrodes connected to a wire and on coupling to a dissipative
environment \cite{Buchler,Hoyos,Fu}.

Experimentally, there are at least two unusual effects that cannot
be explained by local physics. One of them is the anti-proximity
effect in Zn \cite{Tian} and Al \cite{Singh} nanowires. The other is
superconductor--insulator transition (SIT) in short MoGe nanowires
\cite{Bezryadin,Bollinger} (length 30-300 nm), which is claimed to
be controlled by the \textit{normal state resistance} of a wire with
separatrix set by $R_Q=6.45$ k$\Omega$. No QPS was detected in this
work. Surprisingly, experiments on longer MoGe wires \cite{Lau} did
not reveal the SIT; instead a crossover between superconducting and
insulting variations was observed and interpreted in terms of the
increasing rate of unbound QPS. It was suggested that the
discrepancy between the behavior of short and long MoGe nanowires
indicates the existence of a characteristic wire length separating
the SIT and crossover regimes \cite{BezryadinGoldbart,Meidan}. The
QPS contribution and crossover behavior were also detected in long
PbIn \cite{Giordano}, Nb \cite{Heath} and Al \cite{Chang,Arutyunov}
nanowires.

Experimental studies of 1D superconductors face a technical
challenge of fabricating ultra-narrow \textit{homogeneous} wires.
For amorphous MoGe alloys, this was previously  achieved by
deposition of MoGe on top of suspended carbon nanotubes. With this
method, known as the molecular templating technique
\cite{BezryadinGoldbart}, wires with width down to 8 nm  were
fabricated and measured. Disadvantage of the method is that a wire
can not be made sufficiently long; typically, the length is limited
(depending of a type of carbon nanotubes) by 0.3 - 1 $\mu$m.

In the present work, we use an alternative fabrication method --
high resolution electron beam lithography with negative resist
\cite{Namatsu,Yang}. The technique is advanced; in experiments
testing its resolution limit, we were able to fabricate lines with
the width of about 6 nm. Figure 1A shows a scanning electron
microcopy image of one of the nanowires with the width below 10 nm
used in transport measurements. The method does not have the length
limitation. Another advantage is a possibility to make samples with
a true 4-probe geometry as shown in Fig. 1C. In these samples,
current electrodes, voltage probes and a wire are fabricated
simultaneously from the same original MoGe films. The procedure
produces smooth connections between electrodes and wires as shown in
Figs. 1A and 1B. We fabricated and studied two series of nanowires
from amorphous alloys with a distinct relative content of Mo and Ge:
Mo$_{78}$Ge$_{22}$ and Mo$_{50}$Ge$_{50}$. Details of nanowire
fabrication and measurements are given in supplementary materials
\cite{Suppl}. Figure 1D schematically shows the cross section of a
sample.

\begin{figure}
\begin{center}
\includegraphics[width = 3.0in]{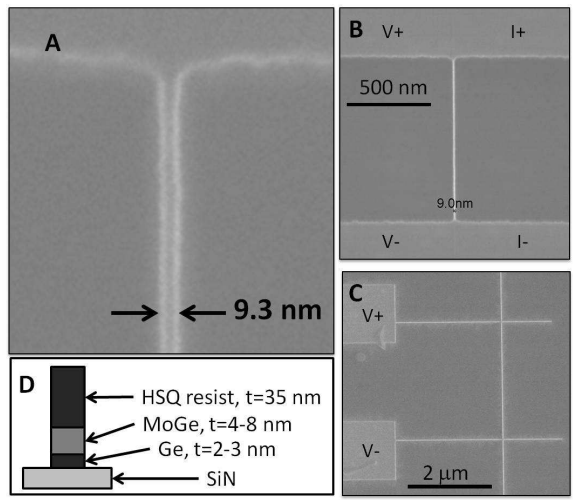}
\caption{\label{fig:nanowires} Sample geometry. (A) Scanning
Electron Microscopy image of a nanowire fabricated by
high-resolution electron beam lithography with negative HSQ resist.
Connection to a film electrode is shown on the top of the figure.
(B) Electrodes geometry for quasi-4-probe transport measurements
used in majority of samples. (C) Electrodes geometry for 4-probe
measurements with voltage-probe electrodes made of nanowires.
Current electrodes are outside of the image.(D) Sketch of the cross
section of a typical sample.}
\end{center}
\end{figure}

\begin {table*}
\setlength{\tabcolsep}{3mm}
\begin {center}
\begin {tabular}{c c c c c c c c c}
\hline\hline $ $ &  $t$ (nm)  &  $w$ (nm)   &    $L$ $(\mu$ m)  &
$R_{RT}$ (k$\Omega$) & $R_{LT}$ (k$\Omega$)& $\rho_L(300$ K)
($\Omega$/nm) &
$A$ (nm$^2$) & $T_c$ (K)\\

\hline

A &  7 & 25 &  3  &  33 & 34.5 & 11 & 145 & 3.0  \\
B &  8 & 19 &  2  &  26 & 28   & 13 & 123 & 2.7  \\
C &  7 & 21 &  9  &  125& 132  & 14 & 115 & 2.7  \\
E &  7 & 20 &  3  &  59 & 62   & 20 &  81 & 1.8  \\
F &  7 & 15 &  25 &  630& 680  & 25 &  64 & 1.1  \\
G &  6 & 17 &  2  &  51 & 56   & 26 &  63 & 1.0  \\
F1&  7 & 15 &  25 &  680& 730  & 27 &  59 & 0.9  \\
H &  7 & 14 &  12 &  356& 381  & 30 &  54 & 0.8  \\
I &  6 & 10 &  8  &  251& 282  & 31 &  51 & 0.6  \\
N &  6 & 9  &  1  &  43 & 45   & 43 &  37 & 0.3  \\
J &  4 & 17 &  3  &  90 & 106  & 30 &  53 & 0    \\
K &  4 & 17 &  18 &  650& 740  & 36 &  44 & 0    \\
M &  4 & 15 &  25 & 1150& 1300 & 46 &  35 & 0    \\

 \hline\hline
\end {tabular}
\caption {The experimental parameters characterizing
Mo$_{78}$Ge$_{22}$ nanowires: $t$ -- nominal thickness; $w$ --
nominal width obtained from an SEM image of a wire; $L$ -- length;
$R_{RT}$ -- resistance at room temperature; $R_{LT}$ -- resistance
at low temperature (2-4 K); $\rho_L=R_{RT}/L$ -- resistance per
length at room temperature; $A$ -- effective cross section area
estimated as $A=\rho L / R_{RT}$, where $\rho$=160 $\mu \Omega$cm is
the bulk resistivity of Mo$_{78}$Ge$_{22}$ amorphous alloy; $T_c$ --
critical temperature taken at the midpoint of a superconducting
transition.}
\end {center}
\end {table*}

The parameters of Mo$_{78}$Ge$_{22}$ nanowires are summarized in
Table 1. The resistance of the wires increases by few percent when
temperature decreases from 300 K down to 2-4 K. This behavior is
typical for strongly disordered systems; the gain in resistance is
due to the weak localization and electron-electron interaction
corrections \cite{Altshuler}. The actual cross section area of
nanowires, $A$, is not exactly known. The thickness of a wire is
reduced from its nominal value (typically by 0.5-1 nm) due to
oxidation and etching by the TMAH developer \cite{Suppl}. As shown
in Fig. 1D, a MoGe nanowire is permanently covered by a 35-nm thick
layer of exposed HSQ e-beam resist. However, the sides of a wire are
not protected so the actual width of a wire can be reduced as a
result of oxidation. Fortunately, MoGe films are known to have
constant volume resistivity ($\rho=160$ $\mu\Omega$ cm) down to
thickness of 1 nm \cite{Graybeal}. This property allows us to
estimate cross sectional area as $A=\rho L/R_{RT}$, where $L$ is
wire length and $R_{RT}$ is resistance at $T$=300 K.

\begin{figure}
\begin{center}
\includegraphics[width = 3.0in]{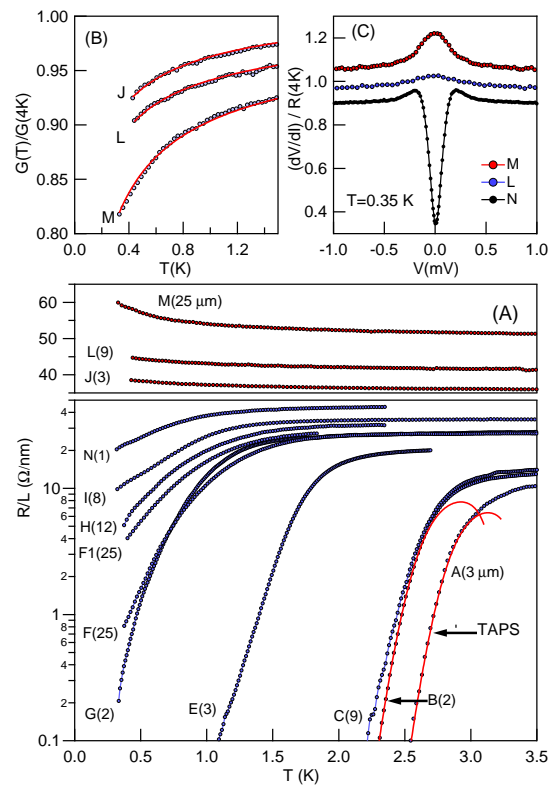}
\caption{\label{fig:nanowires} Superconductor--insulator transition
in zero-magnetic field. (A) Resistance over length versus
temperature for a series of Mo$_{78}$Ge$_{22}$ nanowires. Letters
label the nanowires and numbers in the brackets indicate the
nanowire lengths in micrometers. For wires A and B, solid red lines
are the fitting curves to the theory of thermally activated phase
slips. (B) Low-bias conductance normalized by the value of
conductance at 4 K as a function of temperature for insulating
wires. Solid red lines are the fitting curves to the theory of
electron-electron interactions in 1D. Data for wires L and M are
downshifted by 0.02 and 0.04, respectively. (C) Normalized
differential resistance at $T$=0.35 K for indicated wires. Data for
wires L and N are downshifted by 0.05 and 0.1, respectively.}
\end{center}
\end{figure}

In Fig. 2 we plot resistance per unit length, $\rho_L(T)=R(T)/L$, at
low temperatures for a series of Mo$_{78}$Ge$_{22}$ nanowires. The
wires are labeled by letters and their length in micrometers is
indicated in the parentheses. All wires can be clearly separated in
two groups: superconducting and insulating. Insulating wires
(labeled as M,L,J) have resistance monotonously decreasing with
temperature. As shown in Fig 2B, conductance in these wires can be
well fitted by the dependence $G(T)=G_0-\alpha /\sqrt{T}$ accounting
for the electron-electron interaction in a normal disordered
one-dimension metal (eq. 5.5 in Ref.\cite{Altshuler}). The
one-dimensional approximation is valid when the width of a wire is
smaller than the thermal length $L_T=\sqrt{\hbar D/k_B T}$
\cite{Altshuler}. With the diffusion coefficient of
Mo$_{78}$Ge$_{22}$, $D=0.5$ cm$^2$s$^{-1}$ \cite{Graybeal_PRL}, this
approximation is satisfied for our insulating wires below 1.5 K. The
presence of the SIT is also evident from nonlinear differential
resistance shown in Fig 2C (data were taken at $T$=0.35 K). The
wires always show zero-bias anomaly that changes from negative to
positive when the system crosses the SIT.

The lower panel in the Fig. 2A shows $\rho_L(T)$ for superconducting
wires. We observed that as the resistance per length in the normal
state increases the superconducting transition progressively shifts
to low temperatures. Because the cross sectional area, $A$, of MoGe
nanowires is inversely proportional to $\rho_L(300 K)$ and in the
normal state $\rho_L(T)$ depends on temperature weakly, we can
conclude that superconducting transition in MoGe nanowires is
controlled by $A$. From Table 1 one may notice that the cross
sectional area of insulating wires J and K is larger than that of
superconducting wire N. Wire N has width 9 nm and is the narrowest
wire we have measured. It is possible that this wire has granular
structure as a result of non-uniform side oxidation. However we may
also notice that wire J has nominal cross section 4$\times$17 nm$^2$
and wire N 6$\times$9 nm$^2$. So the alternative explanation is to
assume that the cross sectional area is not the only parameter
controlling the SIT and that for wires with the same $A$, the
superconductivity is stronger when thickness-to-width ratio is
closer to one. Qualitatively similar behavior was observed in wide
MoGe stripes \cite{Graybeal_PRL}.

We do not find any evidence that the length of a wire or its normal
state resistance plays any role in setting superconductivity. This
is further confirmed by measurements on wires A, C, and J that have
4-probe electrode geometry shown in Fig. 1C. We found no difference
in measurements done in quasi-4-probe (Fig. 1B) and 4-probe
geometry, which implies that superconductivity in MoGe wires is not
influenced by electrodes.

Our set of long superconducting wires allows us to make an
interesting observation. The Anderson localization theory predicts
that one-electron states in a disordered normal wire decay
exponentially with a localization length that can be estimated as
$\xi_A=2A k^2_F \ell/3\pi^2$ \cite{Thouless}. Following Ref.
\cite{Graybeal}, we use for a mean free path a value $\ell$=0.3 nm
and estimate the Fermi vector $k_F$ from the free-electron equation
for volume resistivity, $1/\rho=(e^2k_F^2\ell)/(3\pi^2\hbar)$, which
gives $k_F$=1.6 ${\AA}^{-1}$. With these parameters we estimate for
wire F that $\xi_A \approx $300 nm, which is much smaller than the
length of the wire. Therefore, wire F (as well as several other
wires) can be identified as a $\textit{localized superconductor}$.
The term, introduced by Ma and Lee \cite{MaLee}, describes a system
in which superconducting pairing occurs between time-reversed
localized one-electron states. Our observations suggest that the
long-range behavior of one-electron wave functions is not important
for setting superconductivity in long disordered nanowires.

In a 1D superconductor, the finite width of the superconducting
transition arises due to phase slips. For our thickest wires A,B and
C, the $\rho_L(T)$ dependence can be well-explained by the theory of
thermally-activated phase slips (TAPS), as shown in Fig. 2A (we
followed the fitting procedure given in Refs.\cite{Lau} and
\cite{Rogachev1}). Fitting parameters ($T_c$ and the
zero-temperature Ginzburg-Landau coherence length $\xi(0)$) are
$T_c$=3.4 K, $\xi(0)$=9 nm for wire A and $T_c$=3.2 K, $\xi(0)$=9 nm
for wire B. For wires with smaller cross sectional area, fitting
with the TAPS theory is not satisfactory and returns unreasonably
high values of $T_c$ and $\xi(0)$. This trend, observed also in
short MoGe \cite{Rogachev2} and Nb \cite{Rogachev3} nanowires,
possibly reflects a proximity to a zero-temperature quantum phase
transition, where $\xi(0)$ is expected to diverge. The deviation
from the TAPS behavior could be due to an additional contribution
from unbound quantum phase slips. However, for all our
superconducting wires we observe a single-step transition without
any characteristic features of QPS process such as a ``tail'' or
positive curvature in $\rho_L(T)$ curves, or saturation to a
constant resistivity at low temperatures. Our data are markedly
different from the results on series of MoGe nanowires reported by
Lau \textit{et al.}\cite{Lau}, where the QPS contribution was used
to explain behavior of $\rho_L(T)$ dependence.

\begin{figure}
\begin{center}
\includegraphics[width = 3.0in]{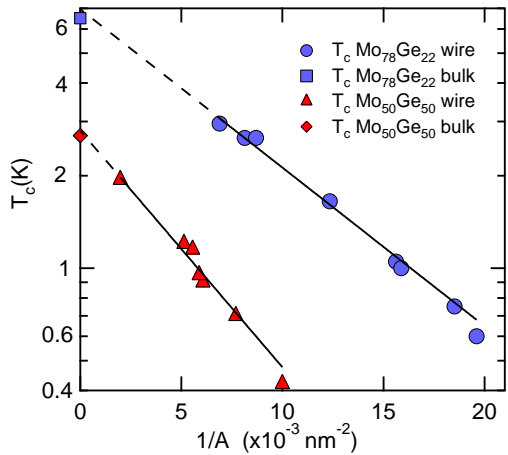}
\caption{\label{fig:nanowires} Critical temperature of
Mo$_{78}$Ge$_{22}$ and Mo$_{50}$Ge$_{50}$ nanowires as a function of
the inverse of the wire cross sectional area. Solid lines represent
a simple exponential form. The dashed lines indicate the extension
of the exponential dependence to the $1/A=0$. Squares indicate the
critical temperature of corresponding bulk alloys.}
\end{center}
\end{figure}

We can also compare our results with the large set of data on short
Mo$_{78}$Ge$_{22}$ nanowires fabricated by the molecular template
technique with length in the range 30-300 nm reported by Bollinger
\textit{et al.} \cite{Bollinger}. All wires in this set do not show
evidence for the QPS or crossover behavior; instead, a direct
superconductor--insulator transition was observed with a separatrix
set by total wire resistance equal to  $R_Q=6.45$ k$\Omega$. The
main evidence for this observation comes from wires with length
smaller than 100 nm.  At low temperatures superconductivity in these
wires can be possibly affected by the proximity effect because of
the attached superconducting electrodes. We found that, if nanowires
with length smaller than 100 nm are excluded, the data provided by
Bollinger match well the data for our long nanowires when plotted in
the $\rho_L(T)$ form. Both sets show a progressive shift of $T_c$
with decreasing cross sectional area, approximately the same width
of superconducting transitions and the same critical value of
$\rho_L\approx$ 40 $\Omega$/nm separating insulating and
superconducting regimes. Combined data cover the range from 100 nm
to 25 $\mu$m clearly indicating that there is no length discrepancy
in the behavior of Mo$_{78}$Ge$_{22}$ nanowires.

For each our wire we can define the empirical mean-field critical
temperature, $T_c$, by choosing it in the middle of the transition.
$T_c$  is plotted in Fig. 3 on a logarithmic scale as a function of
the inverse of the wire cross sectional area, $1/A$, for two series
of wires fabricated form amorphous Mo$_{78}$Ge$_{22}$ and
Mo$_{50}$Ge$_{50}$ alloys. For Mo$_{50}$Ge$_{50}$ the cross
sectional area $A$ was computed from $A=\rho L/R_{RT}$, with
$\rho$=235 $\mu \Omega$cm determined in separate measurements on
film samples. Remarkably, the data for both series can be fitted by
a simple exponential dependence $T_c=T_{c0}$exp$(-\beta/A)$, shown
as a solid line in the figure. The fitting parameter $T_{c0}$ agrees
with the indicated bulk critical temperature of the corresponding
alloy. The second fitting parameter is  $\beta=120$ nm$^2$ for
Mo$_{78}$Ge$_{22}$ and $\beta=180$ nm$^2$ for Mo$_{50}$Ge$_{50}$.

Suppression of $T_c$ by disorder-enhanced Coulomb repulsion was
analyzed theoretically for the crossover region from 2D to 1D
\cite{Oreg,Smith}, and the theory was used to explain the behavior
of $T_c$ in Pb stripes \cite{Sharifi}. Similar to our observation
the fermionic theories predict an exponential suppression of $T_c$;
however, there is a big quantitative disagreement with our data. For
example, we observe experimentally that for a Mo$_{78}$Ge$_{22}$
wire A with an estimated thickness 6 nm and sheet resitance
$R_\Box\approx230$ $\Omega$, reduction of the width from infinity
(2D limit) to 25 nm reduces the critical temperature by 45 $\%$,
from 5.5 to 3 K. On the other hand, when we followed numerical
routines given in Ref.\cite{Oreg} we found that essentially no $T_c$
reduction is expected for this wire. The discrepancy can also be
noticed directly from a comparison with the experimental data on the
Pb stripe with width 22 nm that, unlike the MoGe wire, show no
detectable $T_c$ reduction compared to the 2D case \cite{Sharifi}.
The fermionic theories we used for the analysis do not include the
effect of the Coulomb interaction on the single-particle density of
states. Adding this contribution, as it was done previously for MoGe
films \cite{Kirkpatrick}, may perhaps improve an agreement with the
experiment.

An alternative way to explain the strong deviation of $T_c$ from the
fermionic theory is to assume that $\rho_L(T)$ dependencies are
modified and shifted to lower temperatures by \textit{interacting}
quantum phase slips as was suggested theoretically in Refs.
\cite{GZ_PRL} and \cite{Meidan}. These processes, at least in
principle, could also explain the appearance of the zero-bias
anomaly in the insulating wires and functional behavior of
$\rho_L(T)$ in thin superconducting wires, where it deviates from
the TAPS theory but nevertheless has negative curvature. However,
the overall shift in $T_c$ produced by the interacting QPSs is
expected to depend on a wire length. The lack of any dependence of
$T_c$ on the length on MoGe wires brings strong argument against the
relevance of the interacting QPSs to our system.

To clarify this question further we studied the suppression of
superconductivity in thin wires by magnetic field.
 The variation of the resistivity of wire F1 at different magnetic
 fields (applied normal to the  wire and substrate) is shown in Fig. 4A.
Comparison of the data shown in Figs. 2 and 4 indicates that the
evolution of the resistance in the critical regime both as a
function of temperature and voltage bias is qualitatively the same
for transitions driven by the magnetic field or reduction of cross
sectional area. In both cases, the transition from superconducting
to insulating behavior in $\rho_L(T)$  curves is accompanied by the
sign change of the zero-bias anomaly (ZBA) in differential
resistance as shown in Fig. 4C. It is likely that in both cases the
same physics controls the critical regime of the SIT.  For
superconducting wires far from the critical field the narrowing of
the ZBA with increasing magnetic field (shown as a function of
current in Fig. 4B) simply reflects the decrease of the critical
current of a wire. The origin of the ZBA in insulating and
transitional regimes is not well understood.

\begin{figure}
\begin{center}
\includegraphics[width = 3.0in]{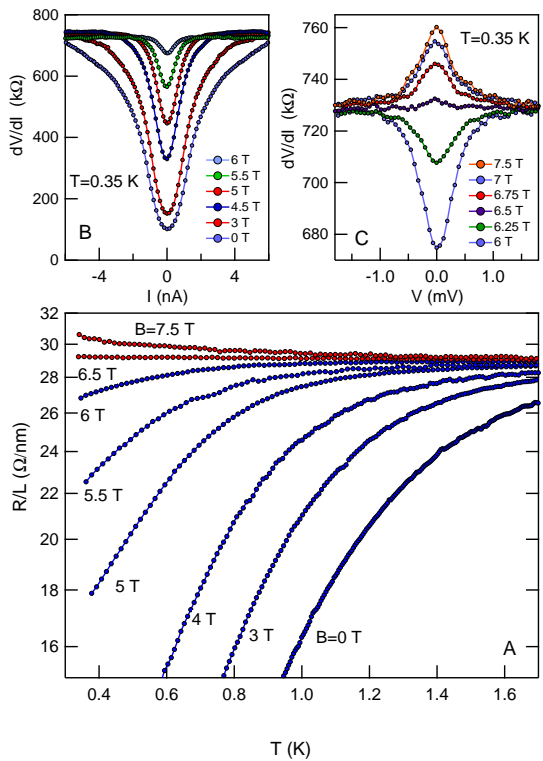}
\caption{\label{fig:nanowires} Superconductor--insulator transition
driven by magnetic field. (A) Temperature dependence of resistance
per unit length for a nanowire F1 at indicated magnetic fields. (B)
Differential resistance as function of current at $T$=0.35 K in
superconducting regime at indicated magnetic fields. (C)
Differential resistance as a function of bias voltage at $T$=0.35 K
in the transitional regime of the SIT.}
\end{center}
\end{figure}

One-dimensional superconductors are too thin to allow formation of
vortexes; instead, magnetic field uniformly penetrates a wire and
suppresses the amplitude of the order parameter acting on the
orbital and spin part of a Cooper pair. The suppression of
superconductivity by a magnetic field in one-dimension is a local
fermionic process. Using parameters of nanowire F1 (mean-field
$T_c=0.9$ K, estimated width $w=10$ nm, diffusion coefficient
$D=0.5$ cm/s$^2$ \cite{Graybeal_PRL}, spin-orbit scattering time
 $\tau_{so}\approx5 \times 10 ^{-14}$ s \cite{Rogachev1}) we can
estimate suppression of $T_c$ from standard formulas for the orbital
($\alpha_o=Dw^2e^2B^2/6\hbar$) and spin
($\alpha_s=\hbar\tau_{so}e^2B^2/2m^2$) pair-breakers \cite{Tinkham}.
Since both contributions are quadratic in a magnetic field we used
the formula $1.76k_BT_c=2\alpha=2(\alpha_o(B_c)+\alpha_s(B_c))$ and
found that $B_c\approx6$ T. It agrees with the experimental value
$B_c\approx6.5$ T. The agreement suggests that the zero-field $T_c$
even for our thinnest wires should be interpreted as a usual
mean-field critical temperature reflecting the reduced magnitude of
the order-parameter.

In summary, we observed the superconductor--insulator transition in
a series of long MoGe nanowires. The SIT, that likely has fermionic
nature, can be driven by wire cross section and by magnetic field.

\begin{acknowledgments}

The authors thank A.M. Finkel'stein, L.B. Ioffe, E.G. Mishchenko, D.
Mozyrsky, Y. Oreg, and  M.E. Raikh for valuable discussions and B.
Baker, M.C. DeLong, and R.C. Polson for technical support. Sample
fabrication was carried out at the University of Utah Microfab and
USTAR facilities. This work is supported by NSF CAREER Grant DMR
0955484.

\end{acknowledgments}

\textbf{Supplementary materials.}

\textbf{1. Nanowire fabrications.} The nanowires were fabricated
using Si wafers covered with a 100 nm layer of SiN and cut in
individual samples with size $6\times9$ mm$^2$. First, using optical
photolithography, consequential deposition of Ti (20 nm) and Au (40
nm) films, and lift-off procedure we fabricated a pattern consisting
of 12 electrodes and several markers. The markers were later used
for alignment and focusing during electron beam lithography. Next,
we sputter deposited a layer of amorphous Ge (thickness 3 nm)
followed (without breaking vacuum) by the sputter deposition of MoGe
alloy (thickness 4-8 nm). The Ge underlayer helped to produce
uniform MoGe films. To make good electrical connection between
pre-patterned Ti/Au electrodes and thin MoGe films, square pads
($5\times5$ micrometers, thickness 30 nm) were placed in each
contact area by positive electron beam lithography with PMAA and
liftoff. After patterning contact pads, the sample was immersed in
the 2.5 $\%$ water solution of TMAH (the developer for negative
electron beam lithography) for clearing. It was realized that TMAH
slightly etches MoGe films with average rate of about 0.2 nm/min. In
the next stage the whole sample was spin coated with 35-nm thick HSQ
(hydrogen silsequioxane) layer. The speciation of the solution is
XR-1541 2 $\%$; it was purchased from Dow Corning. The nanowire and
film electrodes (shown in Fig. 1 of the main text) were patterned by
electron-beam lithography in Nova Nano 630 Scanning Electron
Microscope with standard field-emission Schottky gun and the NPJS
package for beam control. The accelerating voltage was 30 keV; the
dosage was 400-600 $\mu$C/cm$^2$  for area and 3-8 nC/cm for lines.
The exposed pattern was developed in 2.5 $\%$ water solution of TMAH
for 2 min to remove HSQ that was not exposed to the e-beam (negative
lithography). The pattern was etched with reactive ion etching using
SF$_6$ gas. The exposed HSQ resist permanently remains on top of a
wire and electrodes. To prevent oxidation the wires were stored
under vacuum.

\textbf{2. Noise filtering.} Transport measurements were carried out
in a He-3 cryostat. Each line at room temperature was interrupted by
a Pi-filter with cutoff frequency of 1 MHz. The design of
low-temperature filters was copied from the Marcus group at Harvard
University (A.C. Johnson, Appendix C, PhD thesis 2006, Harvard
University). In each line, we added three 100 $\Omega$  resistors
connected in series and tightly anchored into three separate brass
plates. Silver paste was used to seal the resistors in the brass
plate.

\end{document}